\documentclass[10pt,twocolumn,british,tightenlines
,floats,aps,amsmath,amssymb,nofootinbib,superscriptaddress,
prd,showpacs,showkeys,groupedaddress]%
{revtex4-1}
\usepackage{array}
\usepackage{booktabs}
\usepackage{tabu}
\usepackage{dcolumn}
\usepackage{subfigure}
\usepackage{graphicx}
\usepackage{dcolumn}
\usepackage{bm}
\usepackage{slashed}
\usepackage{color}
\usepackage[normalem]{ulem}
\usepackage{babel}
\usepackage{amsmath}
\usepackage{amssymb}
\usepackage{graphicx}
\usepackage{esint}
\usepackage[unicode=true,pdfusetitle,
 bookmarks=true,bookmarksnumbered=false,bookmarksopen=false,
 breaklinks=false,pdfborder={0 0 1},backref=false,colorlinks=false]
 {hyperref}
\usepackage{mathtools}
\makeatletter
\@ifundefined{textcolor}{}
{%
 \definecolor{BLACK}{gray}{0}
 \definecolor{WHITE}{gray}{1}
 \definecolor{RED}{rgb}{1,0,0}
 \definecolor{GREEN}{rgb}{0,1,0}
 \definecolor{BLUE}{rgb}{0,0,1}
 \definecolor{CYAN}{cmyk}{1,0,0,0}
 \definecolor{MAGENTA}{cmyk}{0,1,0,0}
 \definecolor{YELLOW}{cmyk}{0,0,1,0}
}
\usepackage{color}\usepackage{euscript}\usepackage{epsfig}
\usepackage{amsfonts}\usepackage{fancyhdr}
\@ifundefined{textcolor}{}
{%
 \definecolor{BLACK}{gray}{0}
 \definecolor{WHITE}{gray}{1}
 \definecolor{RED}{rgb}{1,0,0}
 \definecolor{GREEN}{rgb}{0,1,0}
 \definecolor{BLUE}{rgb}{0,0,1}
 \definecolor{CYAN}{cmyk}{1,0,0,0}
 \definecolor{MAGENTA}{cmyk}{0,1,0,0}
 \definecolor{YELLOW}{cmyk}{0,0,1,0}
}
\makeatother
\begin{document}
\title{A `Third' Quantization Constructed for Gauge Theory of Gravity}
\author{Maysam Yousefian}
\email{M\_Yousefian@sbu.ac.ir}

\author{Mehrdad Farhoudi }
\email{m-farhoudi@sbu.ac.ir}
\affiliation{Department of Physics,
             Shahid Beheshti University G.C., Evin, Tehran 19839, Iran}
\begin{abstract}
\noindent In general, a global and unique vacuum state cannot be
constructed for a curved space. As a remedy, we introduce a curved
space background geometry with a Minkowski metric tensor and
locally non-zero curvature and torsion. Based on this geometry, we
propose a `third'/vacuum quantization model as a consequence of
Unruh effect. Accordingly, we introduce a  `third' quantization
scalar field as a general coordinate transformation of spacetime
for the second quantization fields. Then we show that in the
classical limit, the `third' quantization fields appear as
Riemannian manifolds with an emergent metric on which the second
quantization fields are located. This way, the standard model of
field theory turns out as  an effective theory. Moreover, using
the proposed `third' quantization fields, we build a $U(1)\times
SU(4)$ Yang-Mills gauge theory for gravity. According to this
gravitational model, we indicate that an analytical solution of
the presented gravitational model, for the `third' quantum field
particle trajectory (such as a star), corresponds to the
trajectory of a test particle in the Mannheim-Kazanas space.
Furthermore, by using non-perturbative methods and lattice gauge
theory results, we render a solution for the potential of the
constructed model that can explain the galaxy rotation curves and
gravitational lensing without any need to dark matter. We also
address the cosmic microwave background  phenomenon and the
expansion of the universe.
\end{abstract}
\date{September 5, 2021}
\maketitle
\section{Introduction}
General relativity is  an elegant and interesting theory of
gravity that is invariant under diffeomorphism  -- see, e.g.,
Refs.~\cite{Janssen2005,Farhoudi2006,FarYou} and references
therein. Physicists have been interested in expressing it (and, in
general, gravitation) as a gauge theory. At the early stages of
gauge theory, in Refs.~\cite{Blagojevic-2013,Utiyama-1956}, a
gauge theory based model of gravity was proposed. Thereafter, by
introducing a gauge theory based on the Poincar\'{e} group for
gravity~\cite{Obukhov-2006,Obukhov-2020}, other types of models
have been suggested but their gauge transformation method differs
compared to the standard approach~\cite{Blagojevic}. Nowadays,
there are many examples of gauge gravitational models built upon
gauge gravitation theory (GGT). Most of these models are developed
by using differential geometry approach -- see
Refs.~\cite{Blagojevic-2013,Utiyama-1956,Obukhov-2006,Obukhov-2020,Blagojevic}
and references therein.

On the other hand, in Ref.~\cite{Lasenby-1998}, a gauge theory of
gravity called gauge theory gravity (GTG) has been constructed in
the language of geometric algebra~\cite{Aragon-1997}. There is an
important difference between GGT and GTG. The background space is
Riemannian for the former, and Minkowski for the latter. The fact
that there is~not a unique vacuum state in a curved background
space~\cite{Wald-2009} is a problem for GGT based models but
not~for GTG. The GTG is the first point of inspiration for the
work presented in this article. The second point has to do with
the second quantization in quantum field theory.

There is a fundamental difference between standard  gauge theories
such as for quantum electrodynamics (QED) and those proposed for
gravity.  Consider a fermionic test particle with mass $m$ and
charge $e$, in a Riemann-Cartan manifold\footnote{The physical
consequences of the existence of torsion have been considered in
the literature. For example, the compatibility of torsion with the
equivalence principle has been discussed in
Ref.~\cite{Heyde-1975}, and the existence of stability in theories
including torsion has been investigated in
Refs.~\cite{Atazadeh-2014,Shabani-2019,Torralba-2020}. Torsion and
discussions about the unitarity, the existence of ghost and some
issues related to quantum field theory have also been discussed in
Refs.~\cite{Shapiro-2000,Shapiro-2002}. However, considering these
types of issues for the proposed model in this study will be
investigated in a separate work.}\
 with connection
$\Gamma^{\alpha}{}_{\mu\nu}$, subjected to an external
electromagnetic gauge field $A^\alpha$. Its autoparallel equation
is
\begin{equation}\label{geo}
 \ddot{x}^\alpha+\Gamma^{\alpha}{}_{\mu\nu}\dot{x}^\mu\dot{x}^\nu
 =\frac{e}{m}(A_{\nu}{}^{,\alpha}-A^\alpha{}_{,\nu})\dot{x}^\nu,
\end{equation}
where the Greek letters run from zero to three. Adding a constant
value to the electromagnetic gauge field will alter the particle's
velocity but, due to Eq.~(\ref{geo}), will keep its acceleration
invariant. Based on QED, a fermion cannot be accelerated by a
quantum of the second quantization gauge connection field. A
statistical set of quanta will be needed\rlap.\footnote{In
Refs.~\cite{Crispino-2008,DeWitt-2014}, it has been shown that an
inertial state is equivalent to an accelerated state on which a
statistical set of the second quantization creation operators is
applied. (Of course, conversely in this work, we will show that an
accelerated state is equivalent to an inertial state on which a
statistical set of the second quantization creation operators is
applied.)}\
 In contrast with the case for gravity,
adding a constant value to the gravitational connection field will
change the particle's acceleration. Inspired by this comparison,
we envisage that a quantum of the gravitational connection field
cannot be a quantum of a second quantization field. Instead, the
effect of it should be similar to the effect of a statistical set
of, or a partition function for, the quanta of the second
quantization gauge connection fields.

For a theory of quantum gravity, we require that the effect of a
quantum of the gravitational connection field be equivalent to
that of a statistical set, or a partition function, of second
quantization quanta. Before defining such a field, one must remedy
the problem of defining a unique vacuum state. For this issue, in
Sec.~II, we first introduce a geometry with a Minkowski background
metric by making changes to the topology of a Riemann-Cartan
manifold (i.e., a manifold with non-zero torsion). In Sec.~III, we
define a field with the required feature as described above. Based
on such a field, we explain that the Unruh
effect~\cite{Crispino-2008} can lead to what can be interpreted as
a `third' quantization. Then, we show that a Rindler vacuum (as a
coordinate transformation of the Minkowski vacuum) can be obtained
by applying a `third' quantization operator (in terms of a
statistical set, or a partition function of, the second
quantization fields) on a Minkowski vacuum.

Based on this approach, we also obtain a general coordinate
transformation of the Minkowski vacuum by applying a `third'
quantization scalar field operator on the Minkowski vacuum.
Subsequently, in Sec.~IV, we indicate that in the classical limit,
the `third' quantization field, regardless of whether it is a
gravitational field or not, plays the role of a Riemannian
manifold on which the second quantization fields are located. In
Sec.~V, we employ the proposed `third' quantization fields to
develop a gauge theory of gravity. Based on this model, in Sec.~VI
we show that some gravitational phenomena at different scales can
be explained without the need for dark matter existence.
Conclusions are drawn in the last section where we also discuss
the phenomenological implications related to the cosmic microwave
background (CMB) and the expansion of the universe.

\section{Minkowski Metric with Non-Zero Curvature and Torsion}
One of the controversial issues in curved space quantum field
theory is about the definition of vacuum state. In general, as the
notion of vacuum state in a curved space is highly
non-unique~\cite{Wald-2009}, defining a global vacuum state for it
is out of reach. However, given that the concept of vacuum in the
Minkowski spacetime is well-defined, if one can somehow replace
the background manifold with it, then it might open a way to
bypass the problem of vacuum definition there.

In the tetrad/frame formalism\rlap,\footnote{Most tensors become
simple in this system, and it has the ability to reflect important
physical aspects of the spacetime however, it does~not alter
reality.}\
 for a tangent bundle, the local relation between the tetradic components of
metric (e.g., $g_{ab}$) and its coordinate ones are
\begin{equation}\label{tetrad}
    g_{ab}=e_{a}{}^{\mu}e_{b}{}^{\nu}g_{\mu\nu},
\end{equation}
where the Latin letters serve to label the tetradic bases. For a
Minkowski frame (i.e., $g_{ab} \equiv \eta_{ab}$), the metricity
$g_{ab;\mu}=0$ makes the spin connection to be
\begin{equation}\label{met-met1}
    \omega^{ab}{}_{\mu}=-\omega^{ba}{}_{\mu}.
\end{equation}
From the topological point of view, the tangent bundle is over the
base manifold $ M $. If one exchanges the fiber with the base
space in the mentioned case, in general, one will have
\begin{equation}\label{met-met2}
    \begin{cases}
        g_{\mu\nu} \equiv \eta_{\mu\nu} \\ g_{\mu\nu;\alpha}\!=\!-\Gamma^{\gamma}{}_{\mu\alpha}
        g_{\gamma\nu}\!-\!\Gamma^{\gamma}{}_{\nu\alpha}g_{\mu\gamma}\!=\!0
    \end{cases}
\!\!\!\!\!\!\!  \rightarrow
    \begin{cases}
        \omega^{ab}{}_{\mu}\neq-\omega^{ba}{}_{\mu}\\\Gamma_{\mu\nu\alpha}=-\Gamma_{\nu\mu\alpha}.
    \end{cases}
\end{equation}
Besides, the connection $\Gamma_{\mu\nu\alpha}$ is just the
contorsion tensor. In this case, the tangent space (as a Minkowski
space with the Greek letters) is considered as the base space, and
the manifold $ M $ (with Latin letters) as the fiber one.

For a local infinitesimal transformation $\Lambda^{a}{}_{b}$
on the manifold $ M $,
\begin{equation}\label{diffeo}
\delta e^{a}{}_{\mu}\!=\!\Lambda^{a}{}_{b}e^{b}{}_{\mu},
\end{equation}
we obtain
\begin{equation}\label{diffeo2}
\delta\omega^{a}{}_{c\mu}
 \!\simeq\!\Lambda^{a}{}_{c,\mu}\!+\!\Lambda^{a}{}_{b}\omega^{b}{}_{c\mu}
 \!-\!\omega^{a}{}_{b\mu}\Lambda^{b}{}_{c}.
\end{equation}
This transformation is a local $GL(4,\mathbb{R})$ gauge
transformation that is isomorphic to local $U(1)\times SU(4)$
gauge transformation. Such a transformation corresponds to
diffeomorphism in the Riemann-Cartan geometry related to
\eqref{met-met1}. We refer to the situation of case
\eqref{met-met2} as a `Minkowski-Cartan' geometry, because the
metric is  Minkowski but with a non-zero torsion. Such a geometry,
in terms of degrees of freedom, corresponds to the Riemann-Cartan
geometry with property \eqref{met-met1}. Accordingly, this
`Minkowski-Cartan' geometry corresponds to a principal $U(1)
\times SU(4)$ bundle, with $\omega^{ab}{}_\mu$ as a connection on
it. $e^a{}_\mu$ is a section of the associated vector bundle, and
the symbol `$;$' is the induced covariant derivative on it.

\section{Quantum Interpretation of Coordinate Transformation via a `Third' Quantization Process}
Inspired by the well-known approach of Ref.~\cite{DeWitt-2014},
from the quantum field theory point of view, we first show that
the Unruh effect can lead to a kind of `third' quantization as
vacuum quantization. For this issue, we represent a Rindler vacuum
as a coordinate transformation of a Minkowski vacuum. This task
will be performed by defining a `third' quantization operator in
terms of a partition function of the second quantization fields,
which acts on the unique Minkowski vacuum constructed in the
previous section. Based on this procedure, we obtain a general
coordinate transformation of the Minkowski vacuum by applying a
`third' quantization scalar field to the Minkowski vacuum. We
interpret this procedure as a `third' (or, vacuum) quantization.
This approach defines a way for representing gravitational
connections in terms of a partition function of the second
quantization fields. Hopefully this should pave a way towards a
quantum process for gravitation.

In this regard, it is also well-known that in an accelerated
reference frame, in the right and left sides of the Rindler wedges
with a uniform proper acceleration (say, $\alpha$) as
\begin{eqnarray}
\begin{cases}
 &\!\!\!\!R: (x,t)=(\alpha^{-1}e^{\alpha\xi}\cosh\alpha\tau,\alpha^{-1}e^{\alpha\xi}\sinh\alpha\tau)\cr
 &\!\!\!\!L: (x,t)=(-\alpha^{-1}e^{\alpha\xi}\cosh\alpha\tau,-\alpha^{-1}e^{\alpha\xi}\sinh\alpha\tau),
\end{cases}
\end{eqnarray}
the Minkowski vacuum appears as a thermal bath, where $(\xi,\tau)$
and $(x,t)$ are respectively the Rindler and Minkowski spacetime
coordinates. Indeed, in Ref.~\cite{DeWitt-2014}, via the
Bogolyubov transformation~\cite{Bogoljubov-1958}, the Minkowski
vacuum has been shown to be equivalent to a gas of the Rindler
quanta in a thermal equilibrium.

Accordingly, in two dimensions, the relation between the Minkowski
vacuum and the Rindler vacuum with a uniform proper acceleration
has been represented as~\cite{DeWitt-2014}
\begin{equation}\label{MiRiVa}
|0_{M}>=\hat{\Psi}_{\alpha}|0_{R,\alpha}>,
\end{equation}
where (in the natural unit $c=1=\hbar$) operator $\hat{\Psi}_{\alpha}$,
in the Rindler space, is
\begin{equation}\label{Psi}
\hat{\Psi}_{\alpha}\equiv\exp{\left[ iW+\sum\limits_{\omega=0}^\infty
e^{-\pi\omega/\alpha} \left(\hat{a}^{R\dagger}_{+\omega}
\hat{a}^{L\dagger}_{+\omega}+\hat{a}^{R\dagger}_{-\omega}
\hat{a}^{L\dagger}_{-\omega}\right)\right] }.
\end{equation}
Here, the term $e^{iW}$ is the vacuum persistence amplitude
with~\cite{DeWitt-2014}
\begin{equation}\label{Vac}
\begin{split}
 &2\,\mathrm{Im}( W)=\ln[\mathrm{tr}\, e^{(-2\pi/\alpha)H}]=(\pi
\alpha/6)\delta(0),\\ \mathrm{Re}&( W)|_{\alpha=0}=0\qquad {\rm
and}\qquad \mathrm{Re}( W)|_{\alpha\neq 0}=\infty .
\end{split}
\end{equation}
The argument of the Dirac delta function is energy,
$\hat{a}^{R\dagger}_{+\omega}$, $\hat{a}^{L \dagger}_{+\omega}$,
$\hat{a}^{R\dagger}_{-\omega}$ and $\hat{a}^{L \dagger}_{-\omega}
$ are respectively the right and left parts of left ($+$) and
right ($-$) moving creation operators (with energy $\omega$) of a
massless scalar field\footnote{For simplicity, we have chosen a
massless scalar field.}\
 $ \hat{\varphi}=\hat{\varphi}_{a}+
\hat{\varphi}_{c} $. The corresponding creation field, in the
Rindler frame, is
\begin{eqnarray}\label{Phi--fi}
\hat{\varphi}_{c}(\xi,\tau)\equiv
 &\!\!\!\!\!\!\!\!\!\!\!\!\!\!\!\!\!\!\!\!\!\!\!\!\!\!\!\!\!\!\!\!\!\!\!\!\int \frac{d\omega}{\sqrt{4\pi\omega}} \left[
  \hat{a}^{R\dagger}_{-\omega}e^{i\omega(\xi-\tau)}\theta(x-t)\right.\cr
 &+\hat{a}^{L\dagger}_{-\omega}e^{i
  \omega(\xi-\tau)}\theta(t-x)+
  \hat{a}^{R\dagger}_{+\omega}e^{i\omega(\xi+\tau)}\theta(x+t)\cr
 &\!\!\!\!\!\!\!\!\!\!\!\!\!\!\!\!\!\!\!\!\!\!\!\!\!\!\!\!\!\!\!\!\!\!\!\!\!\!\!\!\!\!\!\!\!\!\!\!\!\!\!\!
  \left.+\hat{a}^{L\dagger}_{+\omega}e^{i\omega(\xi+\tau)}\theta(-x-t)\right],
\end{eqnarray}
where $\theta(x)$ is the Heaviside step function. The
corresponding annihilation field,
$\hat{\varphi}_{a}=\hat{\varphi}_{c}^{\dagger}$, and the
Hamiltonian is given by
\begin{equation}\label{hami1}
\hat{H}=\sum_{\omega=0}^\infty \omega\left(
\hat{a}^{R\dagger}_{+\omega} \hat{a}^{R}_{+\omega}
+\hat{a}^{R\dagger}_{-\omega}
\hat{a}^{R}_{-\omega}+\hat{a}^{L\dagger}_{+\omega}
\hat{a}^{L}_{+\omega}+\hat{a}^{L\dagger}_{-\omega}
\hat{a}^{L}_{-\omega}\right),
\end{equation}
where the zero-point energy has been omitted. Moreover, the second
quantization independent operators for the Minkowski vacuum (say,
$\hat{b}_{\pm\omega}$ and its Hermitian adjoint) have also been
defined~\cite{Crispino-2008} in terms of the second quantization
operators of the Rindler frame with a uniform proper acceleration,
\begin{equation}\label{BogoTrans}
\begin{pmatrix}
    \hat{b}_{-\omega} \\[0.3em]
    \hat{b}_{+\omega}^{\dagger}
\end{pmatrix}
\!\! =\!\frac{1}{\sqrt{1-e^{-2\pi\omega/\alpha}}}\!\!
\begin{pmatrix}
1 &\!\! -e^{-\pi\omega/\alpha}  \\[0.3em]
-e^{-\pi\omega/\alpha} &\!\! 1
\end{pmatrix}\!\!
\begin{pmatrix}
    \hat{a}^{R}_{+\omega}   \\[0.3em]
    \hat{a}^{L\dagger}_{+\omega}
\end{pmatrix}.
\end{equation}

Now, we intend to find out the inverse relation of \eqref{MiRiVa},
i.e. getting the Rindler vacuum with a uniform proper acceleration
from the Minkowski vacuum. Accordingly, similar to the technique
used in Ref.~\cite{DeWitt-2014} in obtaining relation
\eqref{MiRiVa}, we indicate that the Rindler vacuum can be
achieved in terms of the Minkowski-Fock space in the form of
\begin{equation}\label{RiMiVa}
|0_{R,\alpha}>=\hat{\Phi}_{\alpha}^{\dagger}|0_{M}>,
\end{equation}
with
\begin{equation}\label{Phi}
\hat{\Phi}_{\alpha}^{\dagger}\equiv\exp{\left[
-iW^*-\sum\limits_{\omega=0}^\infty e^{-\pi\omega/\alpha}
\left(\hat{b}^{\dagger}_{+\omega}
\hat{b}^{\dagger}_{-\omega}\right)\right] }
\end{equation}
and the corresponding Hamiltonian
\begin{equation}\label{hami2}
\hat{H}=\sum_{\omega=0}^\infty
\omega\left(\hat{b}^{\dagger}_{+\omega}
\hat{b}_{+\omega}+\hat{b}^{\dagger}_{-\omega}
\hat{b}_{-\omega}\right),
\end{equation}
where again the zero-point energy has been omitted. To prove the
claim of relation \eqref{RiMiVa}, it would be sufficient to show
that the act of any of the corresponding annihilation operators on
this Rindler vacuum vanishes. For this task and without loss of
generality, for instance, we perform the procedure for the
annihilation operator $\hat{a}^{R}_{+\omega}$, wherein, by
transformation \eqref{BogoTrans}, we have
\begin{equation}\label{OpCreat}
\hat{a}^{R}_{+\omega}|0_{R,\alpha}>=\frac{\hat{b}_{-\omega}
    +\hat{b}^{\dagger}_{+\omega}e^{-\pi\omega/\alpha}}{\sqrt{1-e^{-2\pi\omega/\alpha}}}\;
\hat{\Phi}_{\alpha}^{\dagger}|0_{M}>.
\end{equation}
Then, by employing definition \eqref{Phi}, it is straightforward
to show that relation \eqref{OpCreat} vanishes.

On the other hand, in Ref.~\cite{DeWitt-2014}, it has been
indicated that the Minkowski- and Rindler-Fock spaces are
unitarily inequivalent. We furthermore prove that every two
Rindler's vacua with different uniform proper accelerations are
also orthogonal to each other. For this purpose, via relations
\eqref{RiMiVa}, \eqref{Phi} and \eqref{hami2}, we achieve
\begin{eqnarray}\label{RiRiOrth1}
&<0_{R,\alpha'}|0_{R,\alpha}>=<0_{M}|\hat{\Phi}_{\alpha'}\hat{\Phi}_{\alpha}^{\dagger}|0_{M}>=
\cr &
e^{i\mathrm{Re}(W'-W^{*})-\mathrm{Im}(W'+W^{*})}\mathrm{tr}\left[
e^{-\pi(1/\alpha'+1/\alpha)H}\right].
\end{eqnarray}
Then, by using the simple relations
\begin{equation}\label{ExpanTr}
\mathrm{tr}\,
e^{(-2\pi/\alpha)H}=\prod\limits_{\omega=0}^\infty\left[\frac{1}{1-e^{-2\pi\omega/\alpha}}\right]^4
\end{equation}
and
\begin{equation}\label{InEqual}
\left[
\frac{\sqrt{(1-e^{-2\pi\omega/\alpha})(1-e^{-2\pi\omega/\alpha'})}}
{1-e^{-\pi\omega/(1/\alpha+1/\alpha')}}\right]_{\alpha\neq\alpha'}<1,
\end{equation}
and relations~\eqref{Vac}, relation~\eqref{RiRiOrth1} reads
\begin{equation}\label{RiRiOrth2}
<0_{R,\alpha'}|0_{R,\alpha}>=\delta_{\alpha'\alpha}.
\end{equation}
Moreover, due to relation~\eqref{Vac} and definition~\eqref{Phi},
it is clear that
\begin{equation}
<0_{M}|\hat{\Phi}_{\alpha'}^{\dagger}\hat{\Phi}_{\alpha}|0_{M}\!>=e^{i\mathrm{Re}(W'\!-W^{*})-\mathrm{Im}(W'\!+W^{*})}\!=0.
\end{equation}

Therefore, the Minkowski vacuum and the Rindler vacua with
different uniform proper accelerations, not~only are perpendicular
to each other, but each of the latter ones (as `third'
quantization states) can also be obtained via the defined operator
(consisted of a statistical distribution function of the second
quantization operators), which acts on the Minkowski vacuum. Thus,
these vacua form a set of orthogonal bases for their corresponding
Fock space with operators $\hat{\Phi}^{\dagger}_{\alpha}$ and
$\hat{\Phi}_{\alpha}$ as the creation and annihilation operators
with relation
\begin{equation}\label{Com}
<0_{M}|[\hat{\Phi}_{\alpha},\hat{\Phi}_{\alpha'}^{\dagger}]|0_{M}>=\delta_{\alpha\alpha'}.
\end{equation}
Accordingly, we have established a kind of `third' (or, vacuum)
quantization procedure.

As every second quantization operator is usually indexed with a
momentum in a certain direction, these `third' quantization
operators $\hat{\Phi}^{\dagger}_{\alpha}$ and
$\hat{\Phi}_{\alpha}$ have also been indexed with a uniform proper
acceleration in its corresponding direction. Analogously, a vacuum
state, in which each point has a different acceleration, can be
obtained via the act of an operator on the Minkowski vacuum. In
two dimensions, such a Hermitian operator can also be formed from
the Fourier transformation of operators
$\hat{\Phi}^{\dagger}_{\alpha}$ and $\hat{\Phi}_{\alpha}$ as
\begin{equation}\label{DiffAcce}
\hat{\Theta}(x,t)=\sum\limits_{\alpha=-\infty}^{\infty}\left[\hat{\Phi}^{\dagger}_{\alpha}e^{i(\alpha
x-qt)/c^2} +\hat{\Phi}_{\alpha}e^{-i(\alpha x-qt)/c^2}\right],
\end{equation}
where $\alpha/c^2$ and $q/c^2$, in this presented `third'
quantization, are analogous with the wave number and the angular
frequency in the second quantization fields,
respectively\rlap.\footnote{Obviously, the usual unit of $q$ is
$({\rm lenght})^2/({\rm time})^3$.}\
 In this way, we are able to
define a general coordinate transformation of the Minkowski vacuum
using these scalar `third' quantization operators in terms of a
statistical set of the second quantization fields, which can apply
the desired acceleration to a particle of the second quantization
fields. Indeed, this transformation of coordinates deforms
spacetime for the second quantization fields. Since the act of
operator \eqref{DiffAcce} on the Minkowski vacuum produce a scalar
field, such a field does~not cause any torsion or curvature. Of
course, it is clear that the relations of the `third' quantization
fields with each other are similar to the relations of the second
quantization fields with each other. However, in the next section,
while introducing a constant $\mathfrak{h}$ (analogous with the
Dirac constant in the second quantization fields), we indicate
that the states obtained by acting \eqref{DiffAcce} on the
Minkowski vacuum, in the classical limit
$\mathfrak{h}\rightarrow0$, play the role of a Riemannian manifold
on which the second quantization fields are located.

To generalize the presented formulation to four dimensional
spaces, one can simply use the procedure of the quantum field
theory given in Ref.~\cite{Crispino-2008}. In this regard,
consider spaces with coordinates $(x,\mathbf{x}_\perp,t)$ and $
(\xi,\mathbf{x}_\perp,\tau) $ as
\begin{eqnarray}\label{Coord}
\begin{cases}
&\!\!\!\!\!\!\!R:\!\!\!\!\;\;(x,\mathbf{x}_\perp,t)\!\!=\!\!(\alpha^{-1}e^{\alpha\xi}\cosh\alpha\tau,
\mathbf{x}_\perp,\alpha^{-1}e^{\alpha\xi}\sinh\alpha\tau)\cr
&\!\!\!\!\!\!\!L:\!\!\!\!\;\;(x,\mathbf{x}_\perp,t)\!\!=\!\!(-\alpha^{-1}e^{\alpha\xi}\cosh\alpha\tau,
\mathbf{x}_\perp,-\alpha^{-1}e^{\alpha\xi}\sinh\alpha\tau),
\end{cases}\cr&
\end{eqnarray}
where $\xi$ is a spatial dimension in the direction of proper
acceleration and the components of $\mathbf{x}_\perp$ denote the
other two spatial dimensions. Hence, relations \eqref{Phi--fi},
\eqref{BogoTrans} and \eqref{Phi} will respectively change to
\begin{eqnarray}\label{Phi--fi4}
&\hat{\varphi}_{c}(\xi,\tau,\mathbf{x}_\perp)\equiv\int d \omega
\left[
\hat{a}^{R\dagger}_{-\omega\mathbf{k}_\perp}v^{\dagger}_{-\omega\mathbf{k}_\perp}\theta(x-t)\right.
\cr
&+\hat{a}^{L\dagger}_{-\omega\mathbf{k}_\perp}v^{\dagger}_{-\omega\mathbf{k}_\perp}\theta(t-x)+
\hat{a}^{R\dagger}_{+\omega\mathbf{k}_\perp}v^{\dagger}_{+\omega\mathbf{k}_\perp}\theta(x+t)\cr
&\left.
+\hat{a}^{L\dagger}_{+\omega\mathbf{k}_\perp}v^{\dagger}_{+\omega\mathbf{k}_\perp}\theta(-x-t)\right],
\end{eqnarray}
\begin{eqnarray}\label{BogoTrans4}
&\begin{pmatrix}
    \hat{a}^{R}_{+\omega,-\mathbf{k}_\perp}   \\[0.3em]
    \hat{a}^{L\dagger}_{+\omega,+\mathbf{k}_\perp}
\end{pmatrix}
\!\! =\!\frac{1}{\sqrt{1-e^{-2\pi\omega/\alpha}}}\!\!
\begin{pmatrix}
    1 &\!\! e^{-\pi\omega/\alpha}  \\[0.3em]
    e^{-\pi\omega/\alpha} &\!\! 1
\end{pmatrix}\!\!
\begin{pmatrix}
    \hat{b}_{-\omega,-\mathbf{k}_\perp} \\[0.3em]
    \hat{b}_{+\omega,+\mathbf{k}_\perp}^{\dagger}
\end{pmatrix}\cr&
\end{eqnarray}
and
\begin{eqnarray}\label{Phi4}
&\hat{\Phi}_{\alpha}^{\dagger}\!\!\equiv\!\exp{\!\!\left[\!
    -iW^*\!\!-\!\!\sum\limits_{\omega=0}^\infty \!e^{-\pi\omega/\alpha}
    \!\!\left(\!\hat{b}^{\dagger}_{+\omega,+\mathbf{k}_\perp}
    \! \hat{b}^{\dagger}_{-\omega,-\mathbf{k}_\perp}\!\right)\!\right]
    }.
\end{eqnarray}
Here
\begin{eqnarray}
&v_{\pm\omega\mathbf{k}_\perp}\equiv\left[
\frac{\sinh(\pi\omega/\alpha)}{4\pi^4\alpha}\right]^{\frac{1}{2}}K\!_{_{(i\omega/\alpha)}}\!(\frac{|\mathbf{k}_\perp|}{\alpha}
e^{\alpha\xi})\,
e^{i(\mathbf{k}_\perp.\mathbf{x}_\perp\pm\omega\tau)}\cr&
\end{eqnarray}
with $K\!_{_{(\nu)}}\!(\mu)$ as the modified Bessel function, and
$\mathbf{k}_\perp $ as the corresponding wave
vector~\cite{Crispino-2008}.

\section{Relation Between the `Third' and Second Quantization Fields}
Consider a statistical set of particles of the second quantization
fields, which is specified in the form of a partition function
similar to \eqref{Phi} as a scalar `third' quantization field with
the creation and annihilation operators
$(\hat{b}^{\dagger},\hat{b}) $ and the action
\begin{equation}\label{act}
S\propto\sum_{\omega=0}^\infty \left(\!\hat{b}^{\dagger}_{+\omega}
\! \hat{b}^{\dagger}_{-\omega}\!\!\!+\hat{b}_{+\omega}
\!\hat{b}_{-\omega}\!\right).
\end{equation}
If an extra particle of the second quantization fields is added to
this set, action \eqref{act} will slightly alter by $\delta S$.
Given the similarity of $\exp{(-iS/\mathfrak{h})}$ with operator
\eqref{Phi}, the relation between a particle of the second
quantization fields and a scalar `third' quantization field can be
comparable to the relation between $\delta S$ and the pilot wave
$\exp{(-iS/\mathfrak{h})}$ in the `pilot wave theory'\footnote{In
Ref.~\cite{Bohm-1952}, this theory has been used to illustrate the
relation between the first and zeroth quantization or the
classical state.}\
 \cite{Bohm-1952}. To clarify some aspects of
this comparison, let us consider a well-known example in the
acoustic black hole topic as follows.

Analogous with the Gross-Pitaevskii
equation~\cite{Gross-1961,Pitaevskii-1961}, we envisage that the
ground state of a quantum system of identical bosons is described
as\footnote{This equation is a non-relativistic equation with
respect to a maximum speed related to the `third' quantization
fields (say $\mathtt{C}$). In another research \cite{ether}, we
are working on the parameters $\mathfrak{h}$ and $\mathtt{C}$ via
physical constants.}
\begin{equation}\label{NonLinSchor}
\left(
i\mathfrak{h}\partial_{t}+\dfrac{\mathfrak{h}^{2}\boldsymbol{\bigtriangledown}^{2}}{2m}-b|\Theta|^{2}\right)
\Theta(x,t)=0,
\end{equation}
where $\Theta$ is a scalar `third' quantization boson field, $m$
is the mass of field $\Theta$ and $b$ is representative of the
self-interaction power of $\Theta$ with the usual unit $({\rm
lenght})^2/({\rm time})^3$. Afterwards, let us purposely set
$\Theta\equiv\sqrt{\rho}\,\exp{(-iS/\mathfrak{h})}$ and perturb it
as $ \rho\rightarrow\rho_0+\varepsilon\rho_1$ and $S\rightarrow
S_0+ \varepsilon S_1$\rlap.\footnote{Obviously
$\rho=|\Theta|^{2}$, where it can analogously be interpreted as
the corresponding probability density, and $S$ is as the
corresponding action with the unit of $\mathfrak{h}$.}\
 Hence, by substituting these perturbations
into Eq.~\eqref{NonLinSchor}, while using the `pilot wave theory'
and working in the classical limit $\mathfrak{h}\rightarrow 0$
(i.e., neglecting the quantum potential term), in the first
approximation, we achieve the corresponding Hamilton-Jacobi and
the continuity equations, namely
\begin{equation}\label{Pertu1}
\partial_tS_1-\frac{\boldsymbol{p}.\boldsymbol{\bigtriangledown}S_1}{m}-b\rho_1=0
\end{equation}
and
\begin{equation}\label{Pertu2}
\partial_t \rho_1-\boldsymbol{\bigtriangledown}.\left(\rho_0
\boldsymbol{\bigtriangledown} S_1+\rho_1\boldsymbol{p}\right)/m=0.
\end{equation}
Then, by substituting Eq. \eqref{Pertu1} into Eq. \eqref{Pertu2},
it gives\footnote{The $i$ and $j$ indices are assumed to run from
$1$ to $3$.}
\begin{equation}\label{Rieman}
\partial_\mu\left(\sqrt{-g}g^{\mu\nu}\partial_\nu S_1\right)=0,
\end{equation}
where the emerged metric is
\begin{equation}\label{Metric}
\sqrt{-g}g^{\mu\nu} \equiv b^{-1} \begin{pmatrix}
       1 & -p^i/m  \\[0.3em]
       -p^j/m & \quad -(\rho_0b/m)\delta^{ij}+p^ip^j/m^2
     \end{pmatrix}
\end{equation}
and $\boldsymbol{p}=\boldsymbol{\bigtriangledown} S_0$. This
result reveals that a perturbation in a scalar `third'
quantization field (e.g., field $\Theta$), as a second
quantization field, is located on the emergent of a Riemannian
manifold\rlap.\footnote{In
Refs.~\cite{Ge-2010,Barcelo-2011,Ge-2012,Ge-2019}, instead of
employing the Gross-Pitaevskii equation, other kind of equations
have been used. However, their final results are similar to those
obtained in this section.}\
 The
maximum possible speed (i.e., the usual $c$) of such second
quantization field turned out to be $c^2\equiv\rho_0b/m$, as the
realization of interaction properties of a scalar `third'
quantization field. This issue is similar to the sound
subject\rlap,\footnote{In Ref.~\cite{Yousefian}, we have examined
similarities between the elastic waves and the second quantization
fields.}\
 in which the speed of sound is
the realization of the interaction properties of the second
quantization fields~\cite{Haas-2003}. Indeed, the properties (such
as temperature and mass) of the second quantization fields
constitute the properties of sound propagation environment.

Also, it is known that the properties of elastic environments and
their sound waves (as perturbation of the second quantization
fields) are the realization of the properties of the second
quantization fields in a scale of energy. Accordingly, while
considering the last two sections, analogous with the sound waves
and through the presented `third' quantization point of view, we
envisage that the standard model of particle physics and its
parameters would emerge from the properties of the `third'
quantization fields in the corresponding scale of
energy\rlap.\footnote{It is accepted that those constants that
play a fundamental role in the standard model of particle physics
(such as the fine structure constant) vary with the spatial and
temporal changes of gravitational
fields~\cite{Hees-2020,Wilczynska-2020,Hu-2020,Milakovi-2020,Schmidt-2020,Chakrabarti-2021}.}

\section{Modeling `Third' Quantization Fields}
In Sec.~III, we have introduced a `third' quantization scalar
field, which due to the scalar nature of it does~not cause any
curvature and/or torsion. Now in this section, we intend to make a
gravitational model for the `third' quantization fields in such a
way that it would have the following properties. First, we require
that it possesses the `Minkowski-Cartan' geometry presented in
Sec.~II. Second, due to the success of the Weyl gravity as a
renormalizable gravity theory~\cite{Mannheim-2012} in explaining
solar~\cite{Sultana-2012-1},
galactic~\cite{Mannheim-2012-1,Dutta-2018}, extra-galactic and
cluster scales~\cite{Islam-2019} gravitational phenomena, we want
to have a structure and solutions similar to the Weyl-Cartan
theory.

We also want the model to indicate similarities between a rotating
black hole or a rotating star with fermionic elementary particles.
Indeed, there is a not-so-new hypothesis that black holes and
elementary particles are comparable. In the older viewpoints, e.g.
Refs.~\cite{Einstein-1938,Carter-1968,Burinskii-2008}, efforts
were made to show that elementary particles are some kind of black
holes. However more recently, alternative views have emerged
\cite{Johansson-2019,Maybee-2019,Arkani-2019} that hypothesize
black holes behave like elementary particles via the double copy
theory. Finally, we intend to have a kind of symmetry similar to
the one introduced in Ref.~\cite{Yousefian-2019}, proposing that
small scales in physics can simulate large scales.

In this regard, first there is a correspondence between rotating
black holes and fermionic elementary
particles~\cite{Johansson-2019,Maybee-2019,Arkani-2019,Einstein-1938,Carter-1968,Burinskii-2008},
and second, the rotation wave of microstructures of a granular
medium has a fermionic
behavior~\cite{Close,Burnett,Chan,Deymier,Deymier-2017,Deymier-2018}.
Accordingly, while assuming that stars are microstructures of
cosmic structure as a granular medium, we consider stars similar
to fermionic fields\rlap.\footnote{In Ref.~\cite{Yousefian}, we
have shown that the spin wave of microstructures of an elastic
medium behaves similar to a fermion wave. Hence, the rotation of a
microstructure corresponds to the spin of a fermion particle.
Here, in comparison, we have considered that the spin of stars and
planets correspond to the spin of fermionic particles.}\
 In this way, the similarity between a fermionic fundamental
particle and a star can be guaranteed.
 On the other hand, as stars constitute a statistical set of
particles of the second quantization fields, it is plausible to
assume those as `third' quantization fermionic fields. In
addition, since we have chosen a gauge theory for the model, we
presume that a gauge connection (as a gravitational field) can
change the speed and movement of stars.

At this stage, given that we want a model similar to the Weyl
gravity, by considering the structural similarity between the
(general) Yang-Mills theory and the Weyl-Cartan
gravity~\cite{Ivanov-1982,Wheeler-2014,Attard-2016,Wheeler-2018},
we choose a Lagrangian akin to the Yang-Mills theory based on
symmetry group $ U(1)\times SU(4) $ as
\begin{equation}\label{su3}
 \begin{split}
   \mathcal{L}_{\rm YM}&=\mathtt{C}\,\bar{\psi}_a\left( i\,\mathfrak{h}\gamma^\mu
   D_{\mu}{}^a{}_b-m\mathtt{C}\,\delta^a_b\right) \psi^b  \\& -\frac{1}{4} G_{\mu\nu}{}^{ab}G^{\mu\nu}{}_{ab}
   -\frac{1}{4} F_{\mu\nu}F^{\mu\nu}.
 \end{split}
\end{equation}
In this Lagrangian, $\psi^a$ is a spinor fermionic `third'
quantization field and $\bar{\psi}_a$ is its conjugate (both as
the field of stars, analogous with the quark-like particles) with
completeness relation
\begin{equation}\label{completeness}
\psi^a\bar{\psi}_a/\psi^2=\mathbb{I}/4,
\end{equation}
which
lead to
\begin{equation}\label{eta}
    \bar{\psi}_a\gamma_\mu\gamma_\nu\psi^a/\psi^2=\eta_{\mu\nu},
\end{equation}
where $\psi=\sqrt{\bar{\psi}_a\psi^a}$. Also, $G_{\mu\nu}{}^{ab}$ and $ F_{\mu\nu} $ represent the
curvature or strength field tensors, namely
\begin{equation}\label{gauge}
  \begin{split}
 \!\!\!\!\!\!\!\!\!\! G_{\mu\nu}{}^{ab}\!\!=\!\partial_\mu A_\nu{}^{ab}&\!\!-\partial_\nu A_\mu{}^{ab}\!\!+i
    \frac{g}{\mathfrak{h}\mathtt{C}}\!\left( A_{\nu }{}^{a}{}_c A_\mu{}^{cb}
    \!\!-\! A_{\mu }{}^{a}{}_c A_\nu{}^{cb} \right)\!, \\ \!\!\!\!& F_{\mu\nu}=
    \partial_\mu A_\nu-\partial_\nu A_\mu,
  \end{split}
\end{equation}
where $igA_\mu{}^{ab}/(\mathfrak{h}\mathtt{C})$ is a `third'
quantization connection gauge field with its related $ SU(4) $
group symmetry indices $a$ and $ b $,
$ieA_\mu/(\mathfrak{h}\mathtt{C})$ is a `third' quantization
$U(1)$ vector gauge field, $g$ and $e$ are coupling constants
(i.e., $ SU(4) $ and $ U(1) $ charges, respectively). Besides,
\begin{equation}\label{co-deriva}
D_\mu{}^{ab}=\delta^{ab}\partial_\mu-i(gA_{\mu}{}^{ab}+e\delta^{ab}A_{\mu})/\mathfrak{h}\mathtt{C}
\end{equation}
is the gauge covariant derivative and $\gamma^\mu$'s are the Dirac
matrices. Indeed, this Yang-Mills theory is a principal $U(1)
\times SU(4)$ bundle with
$igA_\mu{}^{ab}/(\mathfrak{h}\mathtt{C})$ and
$ieA_\mu/(\mathfrak{h}\mathtt{C})$ as the connections on it,
$\psi^a$ as a section of an associated spinor bundle and
$\gamma^\mu D_\mu{}^{ab}$ as the induced Dirac operator of the
induced covariant derivative on the associated bundle. Now, if we
make a map between the section $\psi^a$ of the spinor bundle in
this Yang-Mills theory and the section $e^a{}_\mu$ of the vector
bundle in the `Minkowski-Cartan' geometry, we can state that the
presented Yang-Mills theory also has the `Minkowski-Cartan'
geometry.

In this regard, we employ an auxiliary constant fermionic Dirac
spinor (say, $\iota$) and its conjugate, $\bar{\iota}$, with the
conditions
\begin{equation}\label{auxiliary}
\bar{\iota}\,\, \iota=4 \qquad\text{and}\qquad \bar{\iota}\,
\gamma_\alpha\,\iota=0.
\end{equation}
Accordingly, if we consider
\begin{equation}\label{tet-con}
     e^a{}_\mu\equiv\bar{\iota}\,\gamma_\mu \psi^a/\psi
\qquad  \text{and}\qquad e_{a\nu}\equiv
\bar{\psi}_a\gamma_\nu\,\iota/\psi
\end{equation}
respectively as a complex tetrad field and its conjugate, then, by
using \eqref{BogoTrans}, \eqref{completeness}, \eqref{eta} and the
Fierz identities~\cite{Okun}, those will lead to
\begin{equation}\label{key}
    \Re (e^a{}_\mu
    e_{a\nu})=\eta_{\mu\nu}\qquad\text{and}\qquad\bar{\psi}_a\,\iota\,\bar{\iota}\,\psi^a=\bar{\psi}_a\psi^a.
\end{equation}
Hence, the corresponding torsion field turns out to
be\footnote{Given the $U(1)$ gauge symmetry in the presented
model, relation \eqref{tors} under the $U(1)$ gauge transformation
can also be written as
 \begin{equation*}
        T^a{}_{\mu\nu} = \psi^{-1}\bar{\iota}\left( D_\nu{}^a{}_{b} \gamma_\mu-D_\mu{}^a{}_{b}\gamma_\nu\right)
        \psi^b.
 \end{equation*}
 }
\begin{equation}\label{tors}
    T^a{}_{\mu\nu} \equiv D_\nu{}^a{}_{b}e^b{}_\mu -D_\mu{}^a{}_{b}e^b{}_\nu.
\end{equation}
Thus, we have stablished a map between the section $\psi^a$ of the
spinor bundle in the presented Yang-Mills theory and the section
$e^a{}_\mu$ of the vector bundle in the `Minkowski-Cartan'
geometry.

Before we continue, let us highlight that Lagrangian \eqref{su3}
is similar to that mentioned in
Refs.~\cite{Ivanov-1982,Wheeler-2014,Attard-2016} for the
Weyl-Cartan theory. For this purpose, we make a map between the
fields of Lagrangian \eqref{su3} and the fields of the Weyl-Cartan
theory as
\begin{equation}\label{analogyDWC}
    \begin{split}
        i\frac{g}{\mathfrak{h}\mathtt{C}}A_{\mu}{}^{ab} &\rightarrow
        \omega^{ab}{}_\mu,
        \\ i\frac{e}{\mathfrak{h}\mathtt{C}}A_\mu &\rightarrow
        K_\mu,
    \end{split}
\end{equation}
where $\omega^{ab}{}_\mu$ is the spin connection as the Lorentz
gauge group, $K_\mu$ is a Weyl vector as the dilatation gauge.
Thus, the curvature tensors of the corresponding connections can
be defined as
\begin{equation}\label{curvature}
    \begin{split}
        \Omega_{\mu\nu}{}^{ab}&\equiv
        i\frac{g}{\mathfrak{h}\mathtt{C}}G_{\mu\nu}{}^{ab},
        \\ \Omega_{\mu\nu}&\equiv i\frac{e}{\mathfrak{h}\mathtt{C}}F_{\mu\nu}\rightarrow\partial_\mu K_\nu-\partial_\nu
        K_\mu.
    \end{split}
\end{equation}
Accordingly, let us consider the Lagrangian
\begin{equation}\label{SU3}
  \begin{split}
   \mathcal{L}_{\rm YM}&=\dfrac{\mathtt{C}}{3}\left(S_a{}^{\mu\nu}T^a{}_{\mu\nu} +
   \!\frac{m\mathtt{C}}{\mathfrak{h}^2}S_a{}^{\mu\nu}
   S^a{}_{\mu\nu} \right) \\&  +\frac{\mathfrak{h}^2\mathtt{C}^2}{4g^2}
   \Omega_{\mu\nu}{}^{ab}\Omega^{\mu\nu}{}_{ab} +\frac{\mathfrak{h}^2
   \mathtt{C}^2}{4e^2} \Omega_{\mu\nu}\Omega^{\mu\nu},
  \end{split}
\end{equation}
where $S_a{}^{\mu\nu}=\psi
\bar{\psi}_a\gamma_\alpha\hat{S}^{\alpha\mu\nu}\iota$,
$S^a{}_{\mu\nu}=\bar{\iota}
\gamma^\alpha\hat{S}_{\alpha\mu\nu}\psi^a/\psi$,
$\sigma_{\mu\nu}=i\left[ \gamma_\mu,\gamma_\nu\right]  / 2$ and
$\hat{S}_{\alpha\mu\nu}\equiv
\mathfrak{h}\{\gamma_\alpha,\sigma_{\mu\nu}\}/8$ is the
corresponding spin tensor operator. This Lagrangian resembles the
one mentioned in Refs.~\cite{Ivanov-1982,Wheeler-2014,Attard-2016}
for the Weyl-Cartan theory\rlap.\footnote{The difference between
Lagrangian \eqref{SU3} and the corresponding one mentioned in
Ref.~\cite{Wheeler-2014} is that in Lagrangian \eqref{SU3} the
spin replaced by the curvature of the special conformal
transformation vector and the square of the spin (as the matter
part) has been added to it. In addition, the symmetric group of
the gravitational connection in Ref.~\cite{Wheeler-2014} is
$SO(1,3)$ while, in Lagrangian \eqref{SU3}, it is $U(1)\times
SU(4)$.}\
 Now, by using relation
\eqref{completeness} and the Dirac equation obtained via the
variation of Lagrangian \eqref{su3} with respect to $\bar{\psi}_a$
and $\psi^a$, and making some manipulations, one obtains
\begin{equation}\label{assum}
    \left(\gamma^\mu D_{\mu}{}^a{}_b\psi^b\right) \bar{\psi}_a=\bar{\psi}_a \gamma^\mu D_{\mu}{}^a{}_b\psi^b\mathbb{I}/4
\end{equation}
and
\begin{equation}\label{assum2}
    \psi^b\left( D_{\mu}{}^a{}_b\bar{\psi}_a\gamma^\mu\right) =\left(D_{\mu}{}^a{}_b \bar{\psi}_a \right) \gamma^\mu \psi^b\mathbb{I}/4.
\end{equation}
Then, by employing relations \eqref{completeness},
\eqref{auxiliary}, \eqref{assum}, \eqref{assum2}, and the Fierz
identities, one can show that \eqref{su3} is the same as
\eqref{SU3}.

In what follows, we show that the equations obtained from the
variation of Lagrangian \eqref{su3} are also related to the ones
obtained from the Einstein-Cartan theory. To perform this task,
first by factoring out $\gamma^\mu$ from the obtained Dirac
equation, we will have\footnote{Note that, every solution of Eq.
\eqref{var-tetr} is also a solution of the obtained Dirac
equation, whereas any solution of the Dirac equation is~not
necessarily a solution of Eq. \eqref{var-tetr}.}
\begin{equation}\label{var-tetr}
     D_\mu{}^a{}_{b}\psi^b=-i\frac{m\mathtt{C}}{4\mathfrak{h}}\gamma_\mu\psi^a.
\end{equation}
Then, by constituting the wedge product of $\gamma_\mu$ and
$D_\mu$ into Eq. \eqref{var-tetr}, and making some manipulation,
one gets
\begin{equation}\label{cartan1}
  \begin{split}
   \left(\right. &\left. D_\nu{}^a{}_{b}\gamma_\mu -D_\mu{}^a{}_{b}\gamma_\nu\right) \psi^b/\psi=
   \frac{m\mathtt{C}}{2\mathfrak{h}}\sigma_{\nu\mu}\psi^a/\psi \\ & =\frac{m\mathtt{C}}
   {8\mathfrak{h}}\{\gamma_\alpha,\sigma_{\nu\mu}\}\gamma^\alpha\psi^a/\psi= \frac{m\mathtt{C}}
   {\mathfrak{h}^2}\hat{S}_{\alpha\nu\mu}\gamma^\alpha\psi^a/\psi
  \end{split}
\end{equation}
and
\begin{equation}\label{cartan2}
 \begin{split}
  \left(\right. &\left. D_\nu{}^a{}_{b} D_\mu{}^b{}_c- D_\mu{}^a{}_{b}D_\nu{}^b{}_c\right)
  \gamma^\nu \psi^c/\psi \\ & =\left( \frac{m\mathtt{C}}{2\mathfrak{h}^2}\right)^2
  \hat{S}^{\alpha\gamma\beta}\hat{S}_{\alpha\gamma\beta}\gamma_\mu
  \psi^a/\psi,
 \end{split}
\end{equation}
where
$\hat{S}^{\alpha\gamma\beta}\hat{S}_{\alpha\gamma\beta}=3\mathfrak{h}^2\mathbb{I}/2$
has been used. By multiplying $\bar{\iota}$ into Eqs.
\eqref{cartan1} and \eqref{cartan2}, while using definitions
\eqref{tet-con} and $S^a{}_{\mu\nu}$, we will have
\begin{equation}\label{cartan01}
        \left(D_\nu{}^a{}_{b}e^b{}_\mu -D_\mu{}^a{}_{b}e^b{}_\nu\right)=
         \frac{m\mathtt{C}}
        {\mathfrak{h}^2}S^a{}_{\mu\nu}
\end{equation}
and
\begin{equation}\label{cartan02}
        \left( D_\nu{}^a{}_{b} D_\mu{}^b{}_c- D_\mu{}^a{}_{b}D_\nu{}^b{}_c\right)
        e^{c\nu}  = \frac{3(m\mathtt{C})^2}{8}
        e^a{}_\mu.
\end{equation}
Now, due to relations \eqref{analogyDWC} and \eqref{curvature},
Eqs. \eqref{cartan01} and \eqref{cartan02} are comparable to the
Einstein-Cartan equations\rlap.\footnote{However, these two
equations are more comprehensive than the Einstein-Cartan
equations. This comprehensiveness is due to the fact that these
two equations also include the Weyl connection. In addition, the
connection in the Einstein-Cartan gravity is a spin connection,
whereas in these equations, the connection is a gauge connection
of the $SU(4)$ symmetry group. Thus, the solutions of the
Einstein-Cartan gravity equations in this particular example can
be regarded as a special case of the solutions of these two
equations.}

Moreover, it is obvious that Eq.~\eqref{cartan02} corresponds to
the Einstein equation with constant curvature (see e.g.,
Ref.~\cite{Shapiro-2002}), and hence some solutions of
Eq.~\eqref{cartan02} for four-velocity of fermionic fields can
correspond to the Schwarzschild-de~Sitter
solution~\cite{DeWitt-1973} for four-velocity of test particles.
Also, as Lagrangian \eqref{su3} and Eqs. \eqref{cartan01} and
\eqref{cartan02} are $U(1)$ gauge invariance, the proposed
Yang-Mills theory results a $U(1)$ transformation of tetrad
$e^a{}_\mu$ (in analogue with a conformal transformation in the
Weyl-Cartan gravity) that is also the solution of Eq.
\eqref{cartan02}. Thus, in addition to the Schwarzschild-de~Sitter
solution, the Mannheim-Kazanas solution~\cite{Mannheim-1989} is
also valid\rlap.\footnote{In Ref.~\cite{Sultana-2017}, it has been
shown that the Schwarzschild-de~Sitter and the Mannheim-Kazanas
solutions are related via a conformal transformation.}\
 With these explanations, the similarity between
the Mannheim-Kazanas solution and the phenomenological potential
between quarks (as represented fields based on the (general)
Yang-Mills theory) is plausible, as pointed in
Ref.~\cite{Mannheim-1989}.

The difference between the presented Yang-Mills theory and the
Weyl-Cartan theory is that the former possesses the
`Minkowski-Cartan' geometry whose base space is the Minkowski
space and its fiber space is a general manifold with $U(1)\times
SU(4)$ symmetry. Whereas, the latter has the Weyl-Cartan geometry
whose base space is a general manifold with torsion and the Weyl
connection, and its fiber space is the Minkowski space.
Topologically, the total space of the presented Yang-Mills theory
corresponds to the total space in the Weyl-Cartan gravity, where
its fiber and base spaces have been displaced.

It is known that one of the major advantages of the (general)
Yang-Mills theory over the Weyl-Cartan theory and other
gravitational theories is that, in general, it is a well-tested
theory and its phenomenological behaviors are well-known, see
e.g., Refs.~\cite{Steinberger-2005,Reimer-2016}. Indeed, the
result of calculations in this theory corresponds to the result of
observations with good accuracy~\cite{Burkert-2018,Shanahan-2019}.
In this respect, in the next section, we use the methods of the
(general) Yang-Mills theory to obtain the effective potential
function for the fields of the presented Yang-Mills theory to
explain the trajectory of stars in the gravitational fields of
galaxies and the gravitational lensing.

\section{Galaxy Rotation Curves and Gravitational Lensing}
In this section, we first address the gravitational potential
obtained from the Yang-Mills theory presented in the previous
section. Due to Eqs. \eqref{cartan01} and \eqref{cartan02}, the
trajectory of a fermion (as a star) in the presented Yang-Mills
theory can correspond to the trajectory of a test particle in the
Weyl-Cartan theory of gravity. That is, for the mentioned
analytical solution, the trajectory of a fermion in the former
theory corresponds to the trajectory of a test particle in the
Mannheim-Kazanas space. In this regard, the static symmetrical
metric of the Mannheim-Kazanas solution (as an analytical solution
of the Weyl conformal gravity) for a point particle is
\begin{equation}\label{effec-poten}
    ds^2=f(r)dt^2-f^{-1}(r)dr^2-r^2(d\theta^2+\sin^2\!\theta\; d\phi^2)
\end{equation}
with the gravitational potential $g_{00}$ as
\begin{equation}\label{g-poten}
    f(r)=V_0-\dfrac{\beta}{r}+\gamma r-\lambda r^2.
\end{equation}
Here $ V_0 $, $ \beta $, $ \gamma $ and $ \lambda $ are some
constants of integration. It has been shown that such a potential
can explain the gravitational phenomena from solar system to
cluster scale without using the concept of dark
matter~\cite{Mannheim-2012,Sultana-2012-1,Mannheim-2012-1,Dutta-2018,Islam-2019}.
However, in a Yang-Mills theory, the screening and confinement
effects are important aspects of the theory\rlap.\footnote{Given
the similarity between the presented Yang-Mills theory and the
Weyl-Cartan one, such effects are expected for the Weyl-Cartan
theory as well.}\
 In what follows, we address the deviation of the effective
potential function from the analytical solution \eqref{g-poten}
and/or the Cornell potential (see \eqref{cornell}).

In the literature, for obtaining a suitable potential for the
(general) Yang-Mills theory, the non-perturbative methods of
Yang-Mills theory (such as lattice field theory) have been
employed. Actually, in the non-relativistic limit (note that in
our case, such a limit should be relative to the speed
$\mathtt{C}$), one possible method is to display the potential as
a Fourier transform of the propagator of the connection field in
the real time formalism~\cite{Thakur-2014,Guo-2019}. Another
method, while using the Wilson
loop~\cite{Rothkopf-2013,Rothkopf-2020} in the lattice field
theory, is to display the potential as
\begin{equation}\label{poten-Wilson}
    V(r)=\lim_{t\longrightarrow \infty}
    \frac{\partial_tW(r,t)}{W(r,t)},
\end{equation}
wherein the Wilson loop is defined as
\begin{equation}\label{Wilson}
    W(r,t)\equiv \left\langle \mathcal{P}\exp\left[ \oint dx^{\mu}A_\mu{}^{ab}\right]
    \right\rangle,
\end{equation}
with $\mathcal{P}$ as the path-ordering operator.

As a few examples (that can also be applied to the presented
theory as a Yang-Mills one), in addition to some interesting
anisotropic solutions~\cite{Thakur-2014}, the real part of some
isotropic potential functions of the non-Abelian Yang-Mills theory
(that has been obtained through the calculations of the methods
mentioned in
Refs.~\cite{Mocsy-2008,Rothkopf-2013,Thakur-2014,Guo-2019,Matsui-1986,Rothkopf-2020,Bala-2020})
are
\begin{equation}\label{poten-inter-quar1}
    \Re \left( V(r)\right)  =C-\dfrac{\alpha}{r}e^{-\mu r}+\dfrac{\sigma}{\mu}(1-e^{-\mu r}),
\end{equation}
\begin{equation}\label{poten-inter-quar2}
    \Re \left( V(r)\right)  =C-\dfrac{\alpha}{r}e^{-\mu r}+\dfrac{2\sigma}{\mu}(1-e^{-\mu r})-\sigma r e^{-\mu r}
\end{equation}
and
\begin{equation}\label{poten-inter-quar3}
    \Re \left( V(r)\right)  =C-\dfrac{\alpha-2\sigma/\mu^2}{r}e^{-\mu r}-\dfrac{2\sigma/\mu^2}{r},
\end{equation}
with temperature dependence parameters $C$, $\alpha$, $\mu$ and
$\sigma$. Besides, the parameter $\mu$ has been known as the Debye
mass, which is the result of the Debye screening
phenomenon~\cite{Matsui-1986}. The Debye mass, at a temperature
below a certain temperature (or at a radius less than a certain
radius), is equal to zero, in which case the potential function of
\eqref{poten-inter-quar1}, \eqref{poten-inter-quar2} and
\eqref{poten-inter-quar3} becomes the so-called Cornell (or
funnel) potential (as one of the most popular potential models)
\begin{equation}\label{cornell}
    V(r)=C'-\dfrac{\alpha'}{r}+\sigma' r.
\end{equation}
However, in the other areas, the potential changes linearly with
temperature~\cite{Mocsy-2008,Rothkopf-2013,Thakur-2014,Guo-2019,Matsui-1986,Rothkopf-2020,Bala-2020}.
Nevertheless, with very small amounts of the Debye mass, the shape
of their potential functions is in the form of \eqref{g-poten}.

The imaginary part of the potential function of
\eqref{poten-inter-quar1}, \eqref{poten-inter-quar2} and
\eqref{poten-inter-quar3} has been attributed to the Landau
damping phenomenon~\cite{Beraudo-2008}, which--while assuming that
the above potentials are valid in the presented Yang-Mills
theory--its qualitative consequences analogously cause formation
or destruction of stars in our case. However, in general, the
value of the imaginary part of the potential is less than the real
part of it. In addition, at short distances, its value is close to
zero and increases with increasing
distance~\cite{Mocsy-2008,Rothkopf-2013,Thakur-2014,Guo-2019,Matsui-1986,Rothkopf-2020,Bala-2020}.
In comparison to the presented theory, such a result indicates
that no star formation would occur near the center of a galaxy,
and it should take place farther away from the center of galaxy.
Actually, it has been observed that the H~II regions (wherein star
formation takes place) are in the arms of spiral galaxies or
around irregular galaxies~\cite{Kennicutt-1988,Kennicutt-1989}. It
is noticeable that these results are~not limited to the $SU(3)$
symmetry group~\cite{Bali-1995,Caselle-2004,Bonati-2020}.

Let us now examine a specific case in almost more details. As
mentioned above, the potential of the presented Yang-Mills fields
behaves like the analytical solution \eqref{g-poten} and/or the
Cornell potential in a small radius. In connection with the
quantitative explanation of the gravitational lensing phenomenon
on a larger scale than clusters (although this potential is
somewhat
successful~\cite{Cattani-2013,O'Brien-2016,Lim-2017,O'Brien-2017}),
in Ref.~\cite{Dutta-2018}, it has been shown that its calculated
value is slightly higher than its observed value. This result
means that as distance increases, the potential function deviates
from the analytical solution \eqref{g-poten} and/or the Cornell
potential. Indeed, it can be considered that, with increasing
distance (and hence with increasing potential), the effect of
vacuum polarization appears and the potential behavior deviates
from the analytical solution \eqref{g-poten} and/or the Cornell
potential. Such an effect of vacuum polarization would become
apparent when the vacuum non-Abelian permittivity value is
different from
one~\cite{Thakur-2014,Rothkopf-2020,Schneider-2002,Ranjan-2007,Agotiya-2009}.
Therefore, for the effective potential in the momentum space, one
makes~\cite{Thakur-2014,Rothkopf-2020,Schneider-2002,Ranjan-2007,Agotiya-2009}
\begin{equation}\label{eff-pot}
    V(k)\:\:\rightarrow\:\:\tilde{V}(k)=\dfrac{V(k)}{\varepsilon(k,T)}
\end{equation}
where $T$ is temperature, $k$ is momentum, $V(k)$ is the
analytical solution of potential, and the non-Abelian permittivity
is~\cite{Schneider-2002,Agotiya-2009}
\begin{equation}\label{pirimit}
    \varepsilon(k,T)=1+\dfrac{\Pi_L(0,k,T)}{k^2},
\end{equation}
wherein $\Pi_L(0,k,T)$, in its simplest case, is the static limit
of the longitudinal connection gauge self-energy, which is equal
to $\mu^2$~\cite{Schneider-2002}. Now, if we assume $V(k)$ being
the Cornell potential \eqref{cornell} in the momentum space (as a
spacial case of solution \eqref{g-poten}), then the Fourier
transform of $\tilde{V}(k)$ in relation \eqref{eff-pot} will be
equal to \eqref{poten-inter-quar3}. On the other hand, in this
case, solution \eqref{poten-inter-quar3} is the same as the
potential used in the scalar-tensor-vector theory
(MOG)~\cite{Moffat-2006}, wherein it has been claimed that this
potential is capable of explaining the gravitational
phenomena~\cite{Moffat-2006-1,Brownstein-2007,Moffat-2009,Moffat-2018,Islam-2020,Moffat-2020,Negrelli-2020}.

An interesting aspect of the presented Yang-Mills theory is that
the connection fields can form a bound state on their own
(something like glueballs in the QCD) without any need for
fermionic particles (stars). Thus, the connection fields in the
presented Yang-Mills theory can be a good alternative to dark
matter. For more explanation, we refer to an example in the
subject of QCD. As an example of a phenomenon described via the
(general) Yang-Mills theory, one can refer to protons. It has
experimentally been determined that each hadron consists of a
number of valence quarks and a large number of seaquarks that are
float in the viscous sap of
gluons~\cite{Steinberger-2005,Reimer-2016}. In a proton as a
hadron, most of the intra-proton pressure is generated by the
field of gluons (as connection fields), and sea quarks have a much
smaller role in generating intra-proton
pressure~\cite{Shanahan-2019}. Given the similarity between the
presented Yang-Mills theory and the (general) Yang-Mills theory,
by qualitatively comparing a galaxy and its stars with a hadron
and sea quarks within it, such a property of intra-proton pressure
would be similar to property of dark matter, which contains the
largest share of mass in a galaxy and causes the gravity rotation
curves.

\section{Conclusions and Outlook}
We have introduced a new geometry with non-zero local curvature
and torsion with a Minkowski metric tensor. This is named
`Minkowski-Cartan' geometry. The tangent bundle over the
Riemannian-Cartan manifold $M$ has been replaced with an $M$
bundle over the Minkowski space. In this way, the total space
has~not changed but the spin connection is~not necessarily
antisymmetric and the connection becomes a contorsion tensor.
Since the base space is Minkowski, then a unique and well-defined
vacuum can be constructed.

We have shown that the quantum of gravitational connection fields
cannot be realized via second quantization. Due to this issue, we
have defined a field in such a way that its quantum is equivalent
to a statistical set or a partition function of second
quantization fields' quanta. This result is then used to show that
the Unruh effect leads to a `third' quantization as vacuum
quantization. In this regard, we have indicated that the Minkowski
vacuum and the Rindler vacua with different uniform proper
accelerations form an orthogonal basis of a Fock space, which are
related to each other by scalar `third' quantization creation and
annihilation operators (i.e., $\hat{\Phi}^{\dagger}_{\alpha}$ and
$\hat{\Phi}_{\alpha}$). In this way, we are able to define a
general coordinate transformation of the Minkowski vacuum via
acting the Fourier transform of $\hat{\Phi}^{\dagger}_{\alpha}$
and $\hat{\Phi}_{\alpha}$ on the Minkowski vacuum. Using the
`pilot wave theory method' and working in the classical limit
(i.e., neglecting the quantum potential term), the `third'
quantization scalar fields play the role of Riemannian manifold on
which the second quantization fields are located.

Given that these `third' quantization scalar fields lack curvature
and torsion, we build an $U(1)\times SU(4)$ Yang-Mills model of
gravity. This model corresponds to the Weyl-Cartan gravity based
on general covariance. An analytical solution for the `third'
quantum field's particle (such as a star) trajectory of the model
corresponds to the trajectory of a test particle in the
Mannheim-Kazanas space and/or the Cornell potential. Solutions for
a potential of the model addressing large scales could be obtained
by using loop corrections and non-perturbative, lattice gauge
theory, results. Essentially, such solutions correspond to
modified gravity models capable of explaining galaxy rotation
curves and gravitational lensing.
 \vskip0.3cm

\textbf{Perspectives on Cosmology---} A `third' quantization field
is represented by a statistical set of the second quantization
fields or a partition function of those. Analogously, the universe
could be represented by a partition function of the `third'
quantization fields with a Lagrangian (of stars and the
gravitational field among those), say $\mathcal{L}_{_{\rm TQF}}$,
such that
\begin{equation}\label{universe}
    \Psi_{\rm universe}\propto\exp\left(-i\int_{t_0}^{t}dt\,\mathcal{L}_{_{\rm
    TQF}}\right).
\end{equation}
In this manner, and since the CMB radiation is a global image of
the statistical universe, then this radiation would also be a
statistical state. Indeed, it is well-known that the CMB intensity
in terms of frequency is equivalent to the chart of a black-body
radiation. On the other hand, the number density of CMB as a
black-body radiation~\cite{Pathria-2011} and the number density of
a gas of the Rindler quanta at thermal
equilibrium~\cite{DeWitt-2014} are quantitatively equivalent. As
mentioned in Sec.~III, the Rindler quanta are partition functions
of the second quantization fields. Therefore, the CMB should also
be a partition function of the second quantization fields (i.e.,
photons). The CMB fluctuations would be indicative of microstates
and microstructures of the `third' quantization fields.

Based on~\eqref{universe}, the universe can either be static for
the case of a completely real $\mathcal{L}_{_{\rm TQF}}$ or
dynamic (expanding or collapsing) for a complex Lagrangian, as
\begin{equation}\label{grow-univ}
    \begin{split}
    &\mathcal{L}_{_{\rm
    TQF}}\longrightarrow\mathcal{L}_{_{\rm
    TQF}}+i\,V_{\rm Im}\\ & \left| \Psi_{\rm universe}\right| \propto\exp\left(\int_{t_0}^{t}dt\,V_{\rm
    Im}\right).
    \end{split}
\end{equation}
The Hubble law confirms an expanding universe at the present era.
Now, as mentioned in Sec.~IV, the `third' quantization fields play
the role of a Riemannian manifold for the second quantization
fields, such as photons. Therefore, the expansion of the universe
implies the expansion of the Riemannian manifold for photons of
the CMB, which leads to a decrease in the CMB temperature.

\section*{Acknowledgements}
We thank the Research Council of Shahid Beheshti University.

%
\end{document}